\documentclass[conference,10pt]{IEEEtran}

\usepackage{graphicx}
\usepackage{amsmath,amssymb}
\usepackage{booktabs}
\usepackage{multirow}

\usepackage{cite}
\usepackage{amsmath,amssymb,amsfonts}
\usepackage{graphicx}
\usepackage{textcomp}
\usepackage{xcolor}
\usepackage{tabularx}

\usepackage{verbatim}
\usepackage{algorithm}
\usepackage[noend]{algorithmic}
\usepackage{etoolbox}
\patchcmd{\maketitle}{\@copyrightspace}{}{}{}
\usepackage{mathptmx} 
\newcommand{\ignore}[1]{}
\usepackage{fancyhdr}
\usepackage[normalem]{ulem}
\usepackage{hyperref}
\usepackage[table]{xcolor}
\usepackage{listings}
\usepackage{tabularx}
\usepackage{setspace}
\usepackage{graphicx}
\usepackage{subcaption}
\usepackage{caption}
\definecolor{codegreen}{rgb}{0,0.6,0}
\definecolor{codegray}{rgb}{0.6,0.6,0.6}
\definecolor{codepurple}{rgb}{0.58,0,0.82}
\definecolor{backcolour}{rgb}{0.99,0.99,0.99}

\lstdefinestyle{mystyle}{
    frame=single,   
    basicstyle=\ttfamily\footnotesize,
    breakatwhitespace=false,         
    breaklines=true,                 
    captionpos=b,                    
    keepspaces=true,                 
    numbers=left,                    
    numbersep=5pt,                  
    showspaces=false,                
    showstringspaces=false,
    showtabs=false,                  
    tabsize=2
}
\newcommand{\ours}{\textsc{GPUFreeLinScaleLLM}}

\title{\ours: Architecting Memory Hierarchy for Scalable LLM Inference on SIMD Multicores}

\author{

\IEEEauthorblockN{Madhumitha Basetti, Aryabartta Sahu}
\IEEEauthorblockA{
Department of CSE\\
IIT Guwahati, India\\
Email: \{m.basetti, asahu\}@iitg.ac.in
}

\and

\IEEEauthorblockN{Vishva Ranjan Singh, Devashree Tripathy }
\IEEEauthorblockA{
Department of CSE\\
IIT Bhubaneswar, India\\
Email: \{a25cs09019,devashreetripathy\}@iitbbs.ac.in
}

}

\date{}

\begin{document}

\maketitle

\begin{abstract}

Matrix-Vector Multiplication (MVM) and Matrix-Matrix Multiplication (MMM) are two critical computational kernels in Large Language Model (LLM) inference, with their relative importance varying across the prefill and decode phases. While prefill processes multiple tokens concurrently and is dominated by compute-intensive MMM, decode repeatedly executes memory-intensive MVM for single-token generation. This work investigates the memory-hierarchy and scalability characteristics of SIMDized LLM inference on multicore processors. 

We propose a memory-hierarchy organization for a SIMD multicore architecture, termed \textit{LinScaleLLM}, in which L1 and L2 caches are private to each core, while the L3 cache and DRAM controller/channel are shared among clusters of four processors. We first characterize SIMDized MVM execution and show that SIMD acceleration reduces arithmetic work while increasing the rate of weight consumption, L2 cache access latency a primary bottleneck to multicore scaling.  To address this bottleneck, we consider private L2 cache per each core and investigate a lightweight, latency-aware private-L2 prefetching strategy that decouples prefetch lookahead from prefetch degree. For regular MVM access streams, deeper lookahead with a low prefetch degree improves prefetch timeliness without substantially increasing DRAM traffic. The D12/K1 (12 cache line look ahead and 1 degree) configuration achieves an L2 demand-miss rate of approximately 3.2--3.4\% up to 16 cores. With improved prefetch timeliness, memory bandwidth becomes the dominant scalability constraint. Provisioning memory-controller resources with increasing core count restores scalability, achieving up to 127.55$\times$ speedup on a 64-core system, with MVM performance comparable to standard and high-end GPUs. On complete llama.cpp decode steps of Llama-3.2-1B in Sniper, the organization generates a token 114.0$\times$ faster on 64 cores than one scalar core without prefetching, and AVX-512 performs within 1.2\% of AVX2. We further characterize SIMDized prefill and show that MMM benefits substantially from data reuse, achieving up to 12.20$\times$ SIMD speedup on a single core and 8.97$\times$ SIMD speedup at eight cores, with an overall 71.23$\times$ speedup over the single-core non-SIMD baseline.

\end{abstract}
\ \\

\section{Introduction}

\label{sec:mvm_mmm_importance}

Matrix-Vector  Multiplication (MVM) and Matrix-Matrix Multiplication (MMM) 
are two of the most critical computational kernels in Large Language Model 
(LLM) inference. However, their relative importance varies significantly 
between the \emph{prefill} and \emph{decode} phases of autoregressive LLM 
inference. In high-end GPU-based server, for prefill-heavy workloads such as long-context RAG and document summarization, prefill can account for approximately 70–90\% of total execution time, whereas decode-heavy workloads, such as long-form code generation, can spend 70–95\% of execution time in the decode phase due to repeated token-by-token generation and memory accesses \cite{DistServe24,Sasathi,WaferLLM}.  
As high-end GPU-based systems require large memory capacities to accommodate increasingly large LLMs, the high cost and limited availability of high-bandwidth memory inside the GPU card are becoming significant challenges for data-center deployments, where as memory cost per GB in CPU based system is relatively low.


During the prefill phase, multiple input tokens are processed concurrently, making matrix--matrix multiplication (MMM), $Y=X\times W$, the dominant computation. Its arithmetic intensity (AI), defined as the ratio of arithmetic operations to data transferred from memory, is approximately $2N/3$ for multiplying two $N\times N$ matrices. Specifically, the computation requires about $2N^3$ operations, while transferring approximately $3N^2$ matrix elements, resulting in high arithmetic intensity and making MMM a compute-intensive kernel. The extensive reuse of matrix elements across computations enables efficient utilization of SIMD/vector units, tensor units, registers, and cache hierarchies, while tiling and blocking further enhance data reuse and reduce memory accesses. Consequently, accelerators optimized for MMM can achieve high prefill throughput, although their performance may be lower during single-token decoding.

MVM is particularly important for autoregressive decoding because the model repeatedly performs operations such as $  \mathbf{y} = W\mathbf{x}$, for every generated token. Typically, only one new token is processed at each decoding step. Large LLM weight matrices must therefore be accessed repeatedly during the generation process.  MVM operations become a dominant component of the decode phase.  Unlike MMM, MVM provides relatively limited opportunities for data reuse. In MMM, the reuse of matrix elements across multiple computations provides significantly higher arithmetic intensity than MVM. Consequently, the MVM  performance is often constrained by the memory subsystem  rather than by the peak computational throughput of the processor.  For a dense MVM involving an $M \times N$ matrix, the arithmetic intensity can be approximated as $ AI \approx   \frac{2MN}{MN + M + N}$. For large weight matrices, the dominant cost is therefore often the movement of model weights from the memory hierarchy. As a result, cache locality,  memory bandwidth, prefetching, SIMD utilization, and parallel execution become critical factors in determining MVM performance.

To characterize this behavior, we profile GPU execution of the \textit{llama.cpp} \cite{touvron2023llama} inference engine using NVIDIA Nsight Compute \cite{nsight_compute} on GPU-based system. The quantized MVM is implemented as \texttt{mul\_mat\_vec\_q}, which is MMVQ kernel. Our analysis shows that  MMVQ kernels dominate decode-stage execution, accounting for up to 94\% of total runtime, while other function consumes less than 6.1\% (as shown in Table \ref{tab:runtime_dist}). Autoregressive decoding invokes many small MMVQ kernels per generated token, which causes frequent CPU–GPU launches. Also, a similar observation is reported in \cite{MLSysSanovarBARR25}, where the decode stage accounts for a substantially larger share of the total execution time than the prefill stage across different prompt-to-output token ratios. Furthermore, in most practical LLM inference scenarios, the input-to-output token ratio is less than one, further emphasizing the dominance of the decode stage in overall inference latency.

\begin{table}[tb!]
\caption{Decode-stage kernel time distribution across GPUs}
\centering
\footnotesize
\setlength{\tabcolsep}{8pt} 
\begin{tabular}{|l|r|r|}
\hline
\textbf{Kernel / Operation} & \textbf{RTX 3060 (12GB)} & \textbf{GTX 740 (4GB)} \ \\
\hline
\cellcolor{yellow}MMVQ (matrix-vector)        & \cellcolor{yellow} \textbf{92.1\%} & \cellcolor{yellow}\textbf{74.4}\% \\ \hline
\cellcolor{yellow}MMVQ-small                  & \cellcolor{yellow}\textbf{2.5\%}  & \cellcolor{yellow}\textbf{20.3\%} \\ \hline
H2D memcpy                  & 1.6\%  & 2.6\% \\ \hline
Memset                      & 2.6\%  & 1.1\% \\ \hline
LayerNorm / RoPE etc.       & 0.5\%  & 0.6\% \\ \hline
\end{tabular}
\label{tab:runtime_dist}
\end{table}

In a CPU-based system, where discrete GPUs are not available, this condition is even worse; the MVM kernel execution time constitutes $\geq$99\% of total inference time.

\begin{table}[tb!]
\centering
\footnotesize
\setlength{\tabcolsep}{3pt} 
\caption{Time share of prefill stage and decode stages for different prompt-to-output token ratios as reported in \cite{MLSysSanovarBARR25}, most of the time, the ratio of input token to output token is much less than 1}
\label{tab:timeshare_prefill_decode}
\begin{tabular}{|c|c|c|c|c|c|c|c|}
\hline
\textbf{Stage} 
& \textbf{64:1} & \textbf{16:1}  & \textbf{8:1}  & \textbf{4:1}  & \textbf{2:1}  & \textbf{1.33:1}  & \textbf{1:1} \\
\hline
Prefill Stage (\%)  & 11.04  & 2.80  & 1.45  & 0.74  & 0.37  & 0.25  & 0.19 \\ \hline
\rowcolor{yellow}
Decode Stage (\%)  & 88.96  & 97.20  & 98.55  & 99.26  & 99.63  & 99.75  & 99.81 \\ 
\hline
\end{tabular}
\end{table}

\begin{figure}[tb!]
    \centering
    \vspace{-0.34cm}
    \begin{minipage}[t]{0.89\columnwidth}
        \centering
        \includegraphics[width=\linewidth]{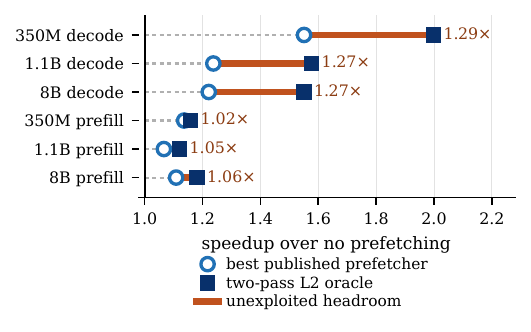}
        \caption*{(a) Headroom}
    \end{minipage}
    \hfill
    \begin{minipage}[t]{0.89\columnwidth}
        \centering
        \includegraphics[width=\linewidth]{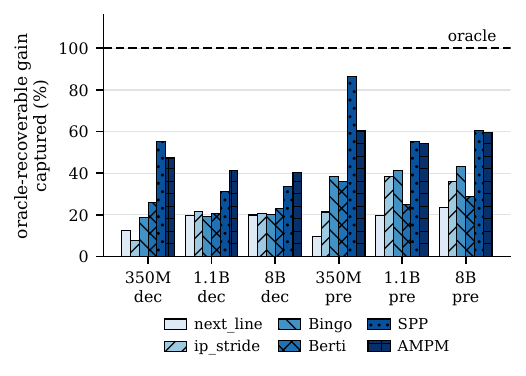}
        \caption*{(b) Oracle capture}
    \end{minipage}
    \vspace{-0.25cm}
    \caption{Prefetching headroom  and oracle-recoverable gain across workloads}
    \label{fig:hr_combined}
\end{figure}

From an architectural perspective, LLM acceleration should therefore consider both MMM and MVM rather than optimizing matrix multiplication as a single homogeneous workload. The prefill phase is generally 
\emph{MMM/GEMM-heavy} and benefits primarily from high compute throughput and 
data reuse, whereas the decode phase is comparatively 
\emph{MVM-heavy} and is often limited by memory bandwidth, cache capacity, 
and data movement. This distinction is particularly relevant for CPU-centric LLM acceleration.  For such systems, improving MVM execution through better cache management,  prefetching, SIMD/AVX utilization, and core-level parallelism can provide substantial improvements in end-to-end LLM generation performance.

Figure \ref{fig:hr_combined} illustrates the prefetching headroom and oracle-recoverable performance gain for LLM inference. Experiments are conducted on memory access traces on ChampSim~\cite{champsim2022} and used all state-of-the-art prefetchers including ip\_stride, SPP, Berti, AMPM, Bingo, etc~\cite{spp2016,bingo2019,berti2022,jouppi1990,bestoffset2016,ipcp2020,ampm}. ChampSim uses a trace-based simulation system with processor core without SIMD. In Fig. \ref{fig:hr_combined}(a), prefill exhibits limited remaining headroom of only 1.02–1.06×, indicating that existing prefetchers already capture most available benefit. Also, the  MMM (prefil) have significantly higher arithmetic density. In contrast, decode shows substantially higher headroom of 1.27–1.29×, revealing considerable untapped prefetching potential. Fig. \ref{fig:hr_combined}(b) further shows that existing prefetchers capture only about 40–55\% of the oracle IPC gain for decode workloads across LFM-350M, TinyLlama-1.1B, and Llama-3-8B.  Here, the oracle is an offline, two-phase replay-like execution, where the oracle knows all the future memory accesses and prefetches before the miss happens. This gap motivates latency-aware lookahead prefetching to recover the remaining performance, particularly for memory-bound decode MVMs. 

These constraints motivate the exploration of cost-effective multicore CPU architectures that can efficiently leverage SIMD parallelism and optimized memory hierarchies for scalable LLM inference.  Also, motivate to a broader question: as SIMD acceleration increases the weight-consumption rate, how should the cache and memory hierarchy be designed to sustain MVM scaling with increasing core count? We show that the performance bottleneck shifts in two stages: from computation to private-L2 prefetch timeliness, and then to off-chip memory bandwidth.

The main contributions of this work are as follows:
\begin{itemize}
    \item \textbf{Memory-Hierarchy Characterization of SIMDized LLM Decoding:}
    We characterize the memory-hierarchy bottlenecks of SIMDized LLM decoding and show that SIMD acceleration substantially reduces arithmetic work while increasing weight-consumption rate in MVM kernels, making \textbf{L2} cache access latency the first bottleneck to multicore scaling.

    \item \textbf{Latency-Aware Private-L2 Prefetching:}
    We consider private L2 cache per each core and investigate a lightweight L2 prefetching strategy that emphasizes prefetch timeliness. For regular MVM access streams, increasing lookahead while maintaining a low prefetch degree improves timeliness without substantially increasing DRAM traffic. D12/K1 achieves a favorable balance, maintaining an L2 demand-miss rate of approximately \textbf{3.2-3.4\%} up to 16 cores.

    \item \textbf{Memory-Bandwidth Provisioning for Multicore Scalability:}
    After alleviating private-L2 latency through timely prefetching, memory bandwidth becomes the dominant scalability constraint. We show that scaling SIMD cores under fixed memory-controller resources provides limited gains, whereas provisioning memory-controller resources with increasing core count restores scalability, achieving up to \textbf{127.55$\times$} speedup on 64 cores system, with performance comparable to MVM execution on standard and high-end GPUs.

    \item \textbf{LLM Prefill on derived SIMDized Multicore:}
    In contrast to decoding, the MMM kernel in prefill benefits substantially from data reuse and compute intensity. Using the same prefetch and architectural configuration as MVM, the SIMD execution of MMM achieves up to \textbf{12.20$\times$} speedup on a single core and \textbf{8.97$\times$} SIMD speedup at eight cores. Combining SIMD and multicore parallelism on octa core system yields an overall \textbf{71.23$\times$ speedup} over the single-core non-SIMD baseline.
        \item\textbf{Evaluation on Real LLM Decoding:}
    Simulating complete llama.cpp decode steps of Llama-3.2-1B in Sniper, we show that the time per token falls with
    the number of memory controllers reaching
    \textbf{114.0$\times$} speedup on 64 cores over one scalar core without prefetching, with AVX2 and AVX-512 within 1.2\% of each other.
\end{itemize}

\section{Comparison of GPU and Multicore AVX Systems for LLM Execution}

The performance of Large Language Models (LLMs) is determined not only by the computational capability of the processing units but also by the ability of the memory subsystem to supply data efficiently.  \textbf{In particular, memory capacity, memory bandwidth, data locality, and data movement are critical factors in determining LLM execution performance}. This creates an interesting opportunity for multicore CPU systems equipped with SIMD extensions such as AVX2 and AVX-512. Although GPUs offer significantly higher parallel computational throughput, multicore CPU systems can provide access to much larger system DRAM capacities at a potentially lower cost per GB. With appropriate memory optimization, SIMD utilization, and data-access scheduling, multicore systems can become viable platforms for executing memory-intensive LLM workloads, particularly when the model size exceeds the available GPU VRAM.  The following diagram illustrates the fundamental architectural differences between the two systems, particularly the organization of compute units, memory hierarchy, and data movement.

\begin{figure*}[tb!]
    \centering
    \includegraphics[width=0.95\linewidth]{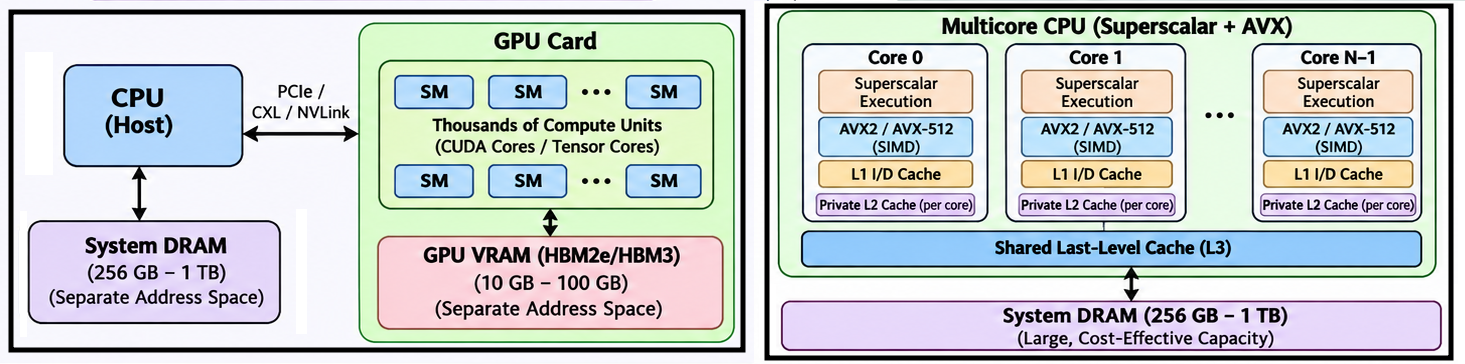}
    \caption{Difference between GPU based system and AVX based multicore system. 
    }
    \label{fig:GPU_CPU_AVX}
\end{figure*}

\ subsection {GPU-based system and its limitations} A GPU-based system consists of thousands of parallel processing units designed to execute computationally intensive operations, particularly matrix multiplication (MMM) and matrix-vector multiplication (MVM). Modern GPUs use dedicated high-bandwidth memory, such as HBM2e, HBM3, or GDDR, physically located on the GPU card. This memory, commonly referred to as GPU VRAM, provides high bandwidth to the GPU's computational units. During LLM execution, model weights are generally loaded into GPU VRAM, allowing the GPU to access them directly without repeatedly transferring them from CPU memory.
 \textbf{An important architectural distinction is that GPU VRAM and CPU system DRAM are two physically separate memory pools with different address spaces in a conventional discrete-GPU system}. CPU system DRAM is connected to the CPU through its memory controllers, whereas GPU VRAM is connected to the GPU's memory controllers. Data movement between these two memory domains typically takes place over PCIe or another high-speed GPU interconnect. Consequently, the available GPU VRAM capacity directly influences the size of the LLM that can be accommodated for efficient execution. Although GPUs provide excellent computational throughput and memory bandwidth, their performance can still become memory-bandwidth-bound, particularly during autoregressive decoding. During decoding, repeated MVM operations fetch large portions of model weights for generating successive tokens. Furthermore, GPU VRAM capacity is relatively limited compared with the system DRAM available in many server configurations, and high-capacity GPU memory configurations can be expensive.

\subsection{Multicore Superscalar CPU with AVX and Shared System Memory}
A multicore superscalar CPU system provides an alternative architecture for LLM execution. In this architecture, multiple CPU cores independently execute instructions using superscalar execution pipelines, while SIMD extensions such as AVX2 and AVX-512 enable each core to perform operations on multiple data elements using a single instruction. AVX2 supports 256-bit vector registers, allowing a single vector operation to process 32 INT8 elements, whereas AVX-512 supports 512-bit vector registers and can process 64 INT8 elements per vector operation. This SIMD capability can accelerate computationally intensive LLM operations, including matrix multiplication and matrix-vector multiplication.
Unlike a discrete-GPU architecture, all CPU cores access a common system memory address space through the CPU memory hierarchy. The system can be equipped with substantially larger DRAM capacities, such as 256 GB, 512 GB, or even 1 TB, depending on the server configuration. This provides an opportunity to accommodate larger LLMs in system memory without requiring the entire model to fit within the comparatively limited GPU VRAM. System DRAM can also offer a lower memory-capacity cost than high-bandwidth GPU memory, although the actual cost depends on the memory technology and system configuration.
However, the availability of large DRAM capacity alone does not guarantee high execution performance. The major challenge is supplying data to the CPU cores at a sufficient rate to keep their SIMD execution units busy. Since LLM inference involves repeated access to large model-weight matrices, inefficient memory access can leave the computational units idle while waiting for data. Therefore, effective memory-access scheduling, hardware and software prefetching, cache utilization, NUMA-aware memory placement, and SIMD optimization become essential for improving performance. In particular, optimizing the timeliness and efficiency of data delivery from DRAM to the CPU caches can help exploit the available memory bandwidth and improve multicore scalability.

\subsection{Central Architectural Observation}

The fundamental difference between these architectures is the trade-off between computational throughput, memory bandwidth, and memory capacity. GPUs provide highly parallel computation supported by dedicated high-bandwidth VRAM, whereas multicore AVX systems provide SIMD-based computation with access to potentially much larger shared system DRAM. For LLM workloads, especially memory-intensive decoding workloads dominated by MVM operations, the ability to efficiently deliver model weights to the processing units is as important as the peak computational capability.  \textbf{Therefore, the key research question is whether an optimized multicore AVX architecture, supported by efficient cache management, latency-aware prefetching, and memory-access scheduling, can exploit large system DRAM to execute LLMs efficiently and achieve competitive performance at a lower memory-capacity cost}.

\section{Related Work}

\subsection{Hardware Data Prefetching}

Hardware data prefetching has been extensively studied to hide memory latency by predicting future cache-line accesses \cite{jouppi1990}. Conventional stream and stride prefetchers exploit regular sequential patterns, while more advanced techniques such as SMS (Spatial Memory Streaming) \cite{SMS} \cite{bingo2019}, VLDP (Variable Length Delta Prefetching), Bingo \cite{bingo2019}, and recent learning-assisted prefetchers improve spatial coverage and prediction accuracy \cite{spp2016, berti2022,fdp2007, bestoffset2016}. These approaches primarily optimize prefetch coverage, accuracy, and bandwidth overhead. Recent studies have also explored adaptive prefetching for irregular and application-specific access patterns. However, prefetch distance and prefetch degree are often treated as coupled parameters, limiting the ability to independently control request timeliness and aggressiveness \cite{spp2016, berti2022,fdp2007, bestoffset2016}. This limitation becomes particularly important for SIMD-accelerated LLM MVM, where wider vector execution increases the cache-line consumption rate and consequently reduces the time available for a prefetched line to reach the target cache before demand access. 

For the regular streaming accesses of quantized MVM, our results show that SIMD acceleration shifts the prefetching challenge from access prediction to timeliness. We therefore evaluate a simple prefetcher that decouples lookahead distance from the number of prefetch requests issued on each demand access. 


\subsection{AVX-512 and OpenMP Parallelization}

SIMD vectorization and thread-level parallelism are widely used to accelerate dense linear-algebra kernels such as MVM and MMM. AVX2 processes multiple low-precision elements per instruction, while AVX-512 provides a 512-bit vector datapath and can process up to 64 INT8 elements simultaneously \cite{AVX512}; AVX-512 VNNI further accelerates INT8 dot-product operations. OpenMP complements SIMD execution by distributing independent matrix rows across multiple CPU cores, typically using static scheduling for regular MVM workloads \cite{pacheco2021introduction}. Recent CPU-based LLM studies demonstrate that vectorization and multithreading substantially improve inference throughput, particularly for quantized models \cite{LLMInferonCPU}. 

\subsection{Memory Access Flow in MVM and MMM}

MVM and MMM exhibit distinct memory-access and data-reuse characteristics that determine the number of active memory flows presented to the memory hierarchy \cite{dao2022flashattention, dao2023flashattention2}. 

For an $M\times K$ INT8 MVM with a 64-byte cache line, each cache line contains 64 weights, resulting in approximately $M\lceil K/64\rceil$ cache-line accesses for a complete matrix traversal \cite{dettmers2022llmint8}. For the $8192\times8192$ MVM studied in this work, this corresponds to 1,048,576 cache lines or approximately 64~MiB of compulsory weight traffic. With $P$ OpenMP threads using contiguous row partitioning, the workload generates approximately $P$ logical sequential weight streams. For each core, there will three flows of request for data, one each for input matrix, input vector and output vector.  In contrast, MMM employs multiple matrix tiles and provides greater temporal reuse of operands, resulting in a different flow structure and lower effective memory pressure depending on blocking. In \cite{MMAccel}, the authors propose StepStone PIM and demonstrate the potential of accelerating MVM operations by performing computation directly in main memory. Their approach exploits underutilized memory bandwidth through processing-in-memory (PIM) units, while avoiding the overhead of moving data between memory and the CPU. 

\subsection{Memory Controller and DRAM Contention}

As SIMD cores are scaled, multiple concurrent memory streams compete for shared DRAM channels, banks, request queues, and memory-controller resources. Prior studies have shown that such contention can increase queuing delays, reduce bank-level parallelism, and limit application scalability even when processor-side cache misses are low \cite{xie2025reimaginingmemoryaccessllm, stagedmemory, kim2012salp}. For memory-intensive workloads, achievable performance therefore depends not only on DRAM bandwidth but also on the number of memory controllers, channel organization, address interleaving, and request scheduling \cite{ChitraSDP25,kim2012salp}. This effect is particularly significant for LLM decoding because large quantized weight matrices generate sustained memory traffic while providing limited reuse during batch-1 MVM.  Aggressive prefetching removes the L2 timeliness bottleneck, increasing core count alone provides diminishing returns under fixed memory bandwidth, whereas increasing memory-controller capacity restores scalability \cite{berti2022, ChitraSDP25}.

\section{Characterization of MVM  on Unicore System}

\begin{table*}[tb!]
\centering
\setlength{\tabcolsep}{2pt}
\caption{System config.  used for evaluation (Snipersim \cite{SniperSim,SniperSimPaper})}
\label{tab:system_configuration}
\renewcommand{\arraystretch}{1.05}
\setlength{\tabcolsep}{2pt}
\begin{tabular}{|l|l|}
\hline
\textbf{Component} & \textbf{Configuration} \\
\hline
Base configs &  nehalem.cfg + gainestown.cfg + rob.cfg \\ \hline
L1 I/D (32KB/4-Way/4Cycle) & Private per core  \\ \hline
L2 (256KB/8-Way/8Cycle)  & Private per core; prefetcher attached at L2 \\ \hline
L3 (8MB/16-Way/30Cycle) & Shared in groups of 4 cores  \\ \hline
DRAM BW and Latency & 45ns base, 7.6GB/s per controller \\ \hline
Controller placement & 1 controller per  4 cores \\ \hline
Native controller counts & 1--4 cores: 1; 8 cores: 2; 16 cores: 4 \\ \hline
\end{tabular}
\end{table*}

\begin{table}[tb!]
\centering
\footnotesize
\caption{Workload structure, MVM with N=8192}
\label{tab:workload_structure}
\setlength{\tabcolsep}{4pt}
\renewcommand{\arraystretch}{1.15}
\begin{tabular}{|l|p{2.1cm}|p{4.0cm}|}
\hline
\textbf{Item} & \textbf{Value} & \textbf{Implication} \\ 
\hline
Matrix &  $NxN$ INT8=64MB &  Streams through memory; much larger than private caches \\ 
\hline
Vector & $N$ INT8=8KB&  Small enough, repeated cache reuse \\  \hline 
Output & $N$ INT32=32KB &  Small relative to matrix \\  \hline
Cache line &  64 B &  Matrix : 1,048,576 cache lines \\  \hline
\end{tabular}
\end{table}

We characterize the MVM Kernel on a single core system (and a multicore system) with SniperSim simulator. The base configuration used for characterization is given in the Table \ref{tab:system_configuration}.

During LLM decoding, quantized matrix-vector multiplication (MVM) repeatedly streams large weight matrices that do not fit in the private caches, while the activation vector is reused across matrix rows. As a result, MVM performance depends on how quickly weight cache lines can be supplied and on the available memory bandwidth. SIMD vectorization reduces the number of arithmetic instructions, but it also causes the core to consume weight cache lines more rapidly, increasing pressure on the memory hierarchy. Because the weight accesses are regular and streaming, effective prefetching primarily requires sufficient lookahead to hide memory-access latency without issuing an unnecessarily large number of prefetch requests. We therefore first study the interaction between SIMD execution and private-L2 prefetching on a single core. This analysis identifies the prefetching requirements of the MVM stream before we examine how increased memory traffic affects multicore scaling under shared off-chip memory bandwidth.

For initial experiment we consider MVM of size 8192x8192 size matrix and 1x8192 vector with each element of 1Byte size. The workload structure is given in the Table \ref{tab:workload_structure}. The matrix footprint alone creates a lower bound of 1,048,576 distinct 64-byte matrix-line fetches on a cold first pass. Therefore, the optimization goal is not to eliminate compulsory matrix traffic; it is to issue those requests early enough that demand accesses do not wait for them, while preserving vector reuse and avoiding additional DRAM traffic.

\subsection{SIMDization of MVM Kernel}

The AVX2 instruction set operates on 256-bit vector registers, enabling simultaneous processing of 32 INT8 elements and substantial data-level parallelism for low-precision workloads. In comparison, AVX-512 doubles the SIMD width to 512 bits, allowing 64 INT8 elements to be processed concurrently. However, unlike AVX-512 VNNI, conventional AVX2 lacks dedicated INT8 dot-product instructions with direct accumulation into 32-bit lanes; consequently, INT8 multiply-accumulate operations require a sequence of unpacking, multiplication, and addition instructions. 

Nevertheless, AVX2/AVX512 provides an effective SIMD execution mechanism for INT8 MVM kernel (also for MMM) kernels commonly used in LLM inference. The single instruction multiple data (SIMD) instructions like AVX2 and AVX-512 reduce arithmetic instruction count and increase throughput, but this also increases cache-line consumption rate. If the hardware prefetcher does not issue memory requests sufficiently early, demand accesses still stall on memory even when the access stream is perfectly regular. Conventional simple prefetchers often couple ``how far ahead" they look with ``how many lines" they issue, creating an inefficient trade-off between insufficient lead time and excessive memory traffic.

The optimized AVX2 kernel processes signed INT8 matrix and vector elements using AVX2 SIMD instructions.
Each 32-byte chunk is widened to 16-bit values, multiplied, pairwise accumulated to 32-bit values, and reduced into the row accumulator. 

OpenMP parallelizes the outer row loop with static scheduling, giving each core contiguous rows.  

\subsection{SIMD Acceleration Exposes the L2 Prefetch-Timeliness Bottleneck}

Sniper's simple prefetcher derives a stride from consecutive accesses and issues num\_prefetches targets using $current\_address + i * stride$. For a 64-byte sequential stream, PF8 reaches at most seven future cache lines (448B) because $i=0$ refers to the current line. This mechanism ties distance to degree: looking farther ahead requires issuing more intermediate requests. That is acceptable when the consumer is slow, but becomes inefficient after SIMD increases the consumption rate.

\begin{lstlisting}
for each qualifying demand access at 
  cache-line address A:
for i = 0 .. K-1:
    prefetch A + (D + i) * 64 bytes   
\end{lstlisting}

Here, $D$ is an independent lookahead distance in cache lines and $K$ is the number of adjacent targets issued at each trigger (D is distance, K is degree \cite{fdp2007}). The selected design is $D12/K1$: target is one cache line prefetch which is $12$ cache lines (768B) ahead of the current demand address. Cross-page operation is enabled for this controlled streaming workload; own-prefetch accesses do not train the prefetcher, and the corrected completion-timing bookkeeping is used. The prefetcher PF8 uses a prefetch degree of 8 (PF8 issues eight targets including the current-line offset, with the farthest target seven cache lines ahead) whereas D12/K1 uses a prefetch distance of 12 cache lines and a prefetch degree of 1.

\begin{table*}[tb!]
\centering
\caption{Single core baseline progression. Scalar and AVX2 rows use
different prefetch settings and should not be interpreted as a pure SIMD
speedup comparison.}
\label{tab:baseline_progression}
\setlength{\tabcolsep}{2pt}
\renewcommand{\arraystretch}{1.15}
\begin{tabular}{|l|c|c|c|c|}
\hline
\textbf{Configuration} & \textbf{ROI time} & \textbf{Instru-} &
\textbf{L2 miss} & \textbf{Total} \\
\textbf{} & \textbf{(ms)} & \textbf{ctions} &
\textbf{} & \textbf{DRAM lines} \\
\hline
Scalar (no prefetcher)  & 106.433 & 469.83 M & 98.28\% & 1,049,770 \\
\hline
Scalar + Simple PF8 & 52.583 & 469.83 M & 6.18\% & 1,049,770 \\
\hline
AVX2 (no prefetcher) & 65.772 & 42.30 M & $\sim$100\% & 1,049,714 \\
\hline
AVX2 + Simple PF8 & 41.206 & 42.30 M & 60.09\% & 1,049,729 \\
\hline
AVX2 + PF8 + SWPF1024 & 37.898 & 48.34 M & 52.39\% & 1,049,824 \\
\hline
AVX2 + Custom D12/K1 & \cellcolor{yellow} 9.052 & $\sim$42--43Ms & \cellcolor{yellow}3.20\% & 1,049,782 \\
\hline
\end{tabular}
\end{table*}

As reported in Table \ref{tab:baseline_progression}. the scalar core with PF8 prefetcher executes about $469.8 M$ instructions, whereas the AVX2 kernel executes about $42.3M$ instructions, an approximately $11.1x$ reduction in instruction count. However, the matrix footprint is unchanged. Consequently, AVX2 consumes cache lines much faster and exposes memory latency.Without hardware prefetching, the L2 miss rate approaches 100\%. Even with PF8, 60.09\% of L2 demand accesses still result in misses.

\begin{table}[tb!]
\centering
\setlength{\tabcolsep}{3pt}
\caption{Performance sensitivity to lookahead depth on single core}
\label{tab:lookahead_sensitivity}
\footnotesize
\renewcommand{\arraystretch}{1.15}
\begin{tabular}{|c|c|c|c|c|c|}
\hline
\textbf{Config} & \textbf{Look} & \textbf{Time} &
\textbf{L2} & \textbf{DRAM} & \textbf{Total} \\
\textbf{} & \textbf{ahead} & \textbf{(ms)} &
\textbf{miss} & \textbf{queue (ns)} & \textbf{DRAM Access} \\
\hline
D4/K1  & 256 B  & 13.162 & \cellcolor{yellow}3.21\% & 0.05   & 1,049,762 \\
\hline
D6/K1  & 384 B  & 9.684  & \cellcolor{yellow}3.20\% & 0.96   & 1,049,762 \\
\hline
D8/K1  & 512 B  & 9.060  & \cellcolor{yellow}3.20\% & 13.25  & 1,049,778 \\
\hline
D12/K1 & 768 B  & 9.052  & \cellcolor{yellow}3.20\% & 46.06  & 1,049,782 \\
\hline
D16/K1 & 1024 B & 9.044  & \cellcolor{yellow}3.20\% & 78.95  & 1,049,802 \\
\hline
D20/K1 & 1280 B & 9.038  & \cellcolor{yellow}3.20\% & 111.82 & 1,049,796 \\
\hline
D24/K1 & 1536 B & 9.027  & \cellcolor{yellow}3.20\% & 144.72 & 1,049,793 \\
\hline
\end{tabular}

\end{table}

\begin{figure}[tb!]
    \centering
    \begin{subfigure}[b]{0.86\linewidth}
        \centering
        \includegraphics[width=\linewidth]{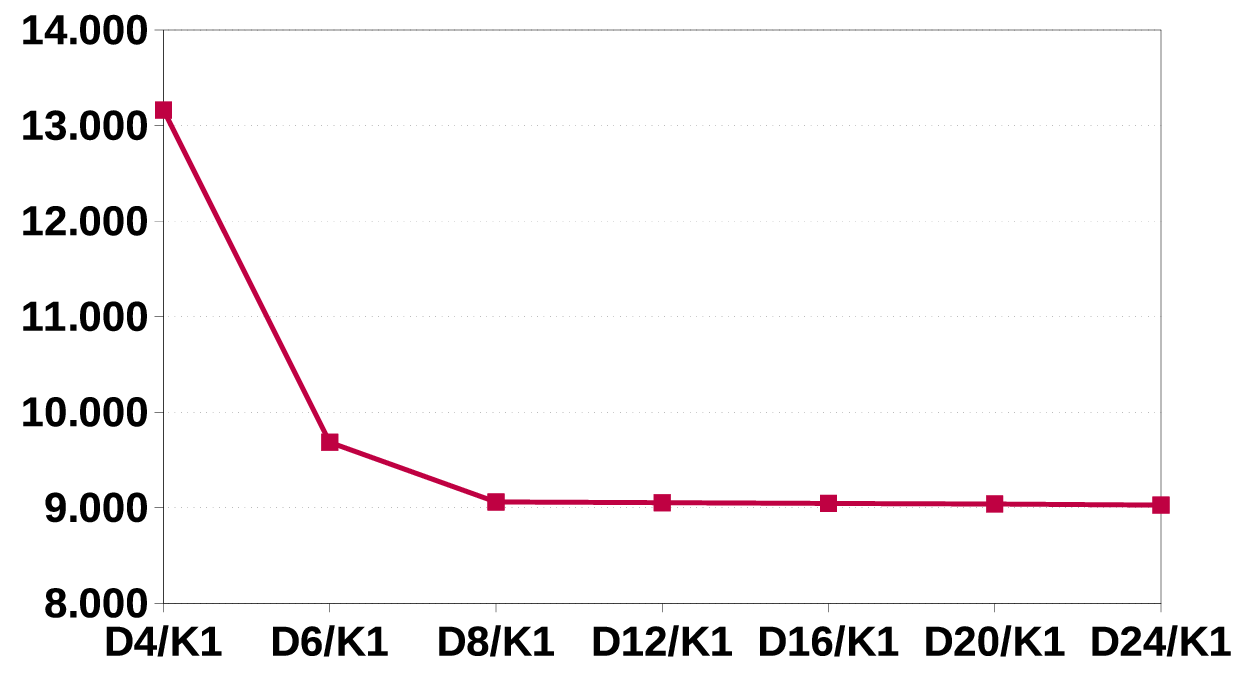}
        \caption{D vs. ROI time}
        \label{fig:D_vs_ROI_time}
    \end{subfigure}
    \hfill
    \begin{subfigure}[b]{0.85\linewidth}
        \centering
        \includegraphics[width=\linewidth]{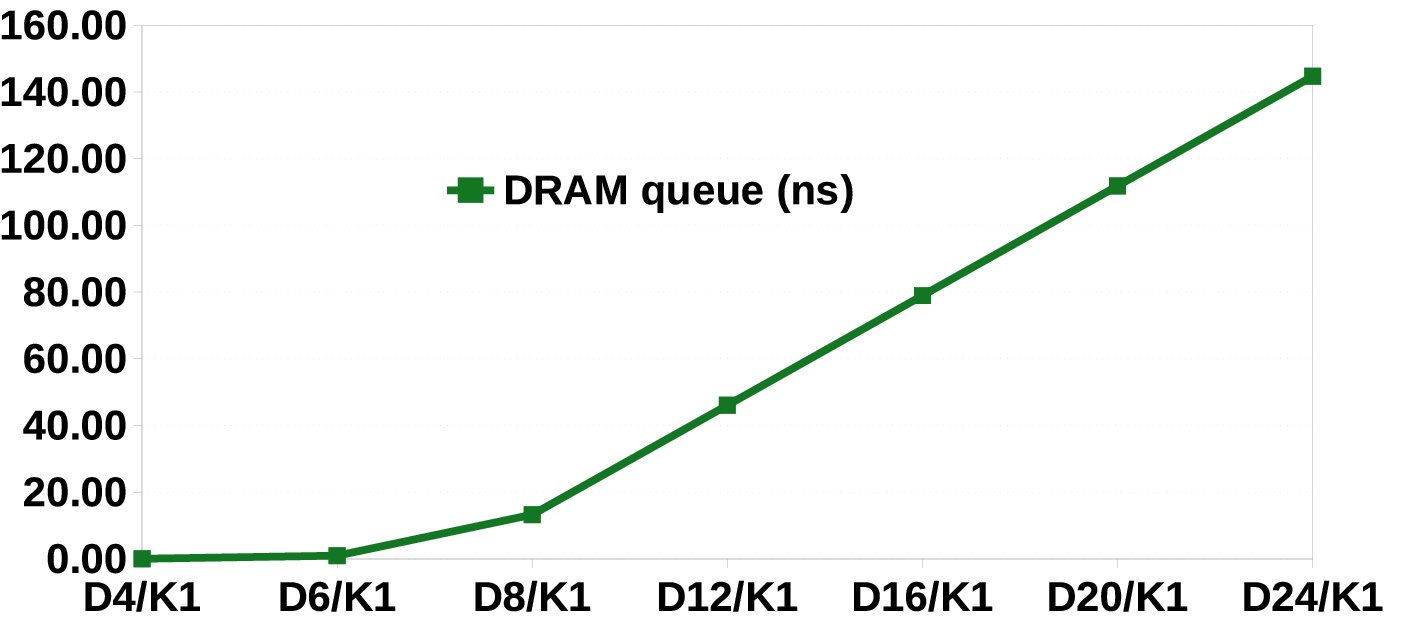}
        \caption{D vs. DRAM queue}
        \label{fig:D_vs_DRAM_Q}
    \end{subfigure}
    \caption{Impact of prefetch distance ($D$) on (a) ROI execution time and 
    (b) DRAM queue latency for a single-core configuration.}
    \label{fig:D_analysis}
\end{figure}

As shown in the Table \ref{tab:lookahead_sensitivity}, the strongest observation is that D4/K1 and D12/K1 both achieve approximately 3.2\% L2 demand-miss rate and roughly the same useful-prefetch coverage, yet D4 takes 13.162 ms while D12 takes 9.052 ms. The difference is not cache coverage but timeliness: at D4, many prefetched lines are requested too late to hide memory latency. By D8, runtime is already 9.060 ms; As reported in Table \ref{tab:lookahead_sensitivity} and shown in Figure \ref{fig:D_vs_DRAM_Q} and Figure \ref{fig:D_vs_ROI_time}, increasing distance through D24 changes runtime by less than about 0.4\%, while DRAM queueing rises from 13.25 ns at D8 to 144.72 ns at D24. Therefore, D8-D12 is the practical knee, and D12 is chosen as a balanced, stable configuration rather than because D12 is universally optimal as can be seen from Figure \ref{fig:D_vs_ROI_time}.

\begin{table}[tb!]
\centering
\footnotesize
\caption{Performance comparison of AVX2 and AVX-512 for different prefetch configurations on a single core. Number of instructions for
AVX2 and AVX512 are 42.3 and 21.32 million, respectively}
\label{tab:avx2_avx512_mvm}
\renewcommand{\arraystretch}{1.05}
\setlength{\tabcolsep}{4pt}
\begin{tabular}{|c|c|c|c|c|c|}
\hline
\textbf{Config} & \textbf{ISA} & \textbf{Time} &
\textbf{L2Miss} & \textbf{L3Miss} & \textbf{BW } \\
&  & \textbf{(ms)} &
\textbf{(\%)} & \textbf{(\%)} & \textbf{(\%)} \\
\hline

No-PF & AVX2 &
65.77 & 100.00 & 96.86 & 13.45 \\
\hline

No-PF & AVX512 &
65.86 & 100.00 & 96.81 & 13.44 \\
\hline

D12/K1 & AVX2 &
9.05 & 3.20 & 96.83 & 97.76 \\
\hline

D12/K1 & AVX512 &
9.05 & 3.25 & 96.78 & 97.77 \\
\hline

D20/K1 & AVX2 &
9.04 & 3.20 & 96.83 & 97.90 \\
\hline

D20/K1 & AVX512 &
9.04 & 3.25 & 96.78 & 97.89 \\
\hline

\end{tabular}
\end{table}

\begin{table}[tb!]
\caption{Performance  across different MVM projection shapes and prefetchers, on a single core with AVX2 (and for  multicore, speedup scales almost linearly)}
\label{tab:mvm_projection_comparison}
\footnotesize
\renewcommand{\arraystretch}{1.05}
\centering
\setlength{\tabcolsep}{5pt}
\begin{tabular}{|c|r|r|r|r|r|}
\hline
\textbf{Matrix} & 
\textbf{Size} & 
\textbf{No PF} & 
\textbf{PF8} & 
\textbf{D12/K1} & 
\textbf{Speedup} \\

\textbf{} & 
\textbf{MiB} & 
\textbf{(ms)} & 
\textbf{(ms)} & 
\textbf{(ms)} & 
\textbf{} \\
\hline
4096 $\times$ 4096   & 16   & 16.457  & 5.600   & 2.242  & \cellcolor{yellow} \textbf{7.34$\times$} \\ \hline
11008 $\times$ 4096 & 43   & 44.262  & 15.073  & 6.045  & \cellcolor{yellow} \textbf{7.32$\times$} \\ \hline
14336 $\times$ 4096 & 56   & 57.572  & 19.531  & 7.792  & \cellcolor{yellow} \textbf{7.39$\times$} \\ \hline
8192 $\times$ 8192   & 64   & 65.772  & 41.206  & 9.052  & \cellcolor{yellow} \textbf{7.27$\times$} \\ \hline
28672 $\times$ 8192 & 224  & 239.889 & 159.520 & 31.156 &\cellcolor{yellow}  \textbf{ 7.70$\times$} \\ \hline

\end{tabular}%
\end{table}

Table \ref{tab:avx2_avx512_mvm} shows the performance comparison of AVX2 and AVX-512 for different prefetch configurations on a single core. Number of instructions for
AVX2 and AVX512 are 42.3 and 21.32 million, respectively. But the execution time and L2 cache miss are not reduced by using AVX512. Table \ref{tab:mvm_projection_comparison} shows performance comparison across different MVM projection shapes and different prefetchers on a single core with AVX2.  For all the MVM sizes, we achieve 7.27$\times$ to 7.70$\times$ speedup on a single-core machine.

\section{MVM  on Multi-Core System}

Similar to single core system, we characterize MVM Kernel on multicore system with SniperSim simulator. The base configuration used for characterization is given in the Table \ref{tab:system_configuration}. 

We explicitly studies this interaction, combining SIMD/OpenMP MVM parallelism with latency-aware prefetching to determine the prefetch distance required to sustain high-throughput execution using AVX instructions. 

For an $M\times K$ INT8 MVM with a 64-byte cache line, each cache line contains 64 weights, resulting in approximately $M\lceil K/64\rceil$ cache-line accesses for a complete matrix traversal \cite{dettmers2022llmint8}. For the $8192\times8192$ MVM considered  in this work, this corresponds to 1,048,576 cache lines or approximately 64~MiB of compulsory weight traffic. In this case, we use simple row wise partitioning of work among processor.  With $P$ OpenMP threads using contiguous row partitioning, the workload generates approximately $P$ logical sequential weight streams. All the cores work independently on different area independently and it can improve the performance linearly with number of cores if the required data is supplied from memory to core efficiently. 

\begin{lstlisting}
void mvm_openmp(const int8_t *A, const int8_t *x, 
    int32_t *y,  int rows, int cols){
#pragma omp parallel for schedule(static)
for (int i=0; i<rows; i++) {
    int32_t sum = 0.0f;
    for (int j=0; j<cols; j++) 
        sum += (int32_t) (A[i*cols + j] * x[j]);        
    y[i] = sum;
    }
} 
\end{lstlisting}

\begin{figure}[tb!]
    \centering
    \includegraphics[width=1.0\linewidth]{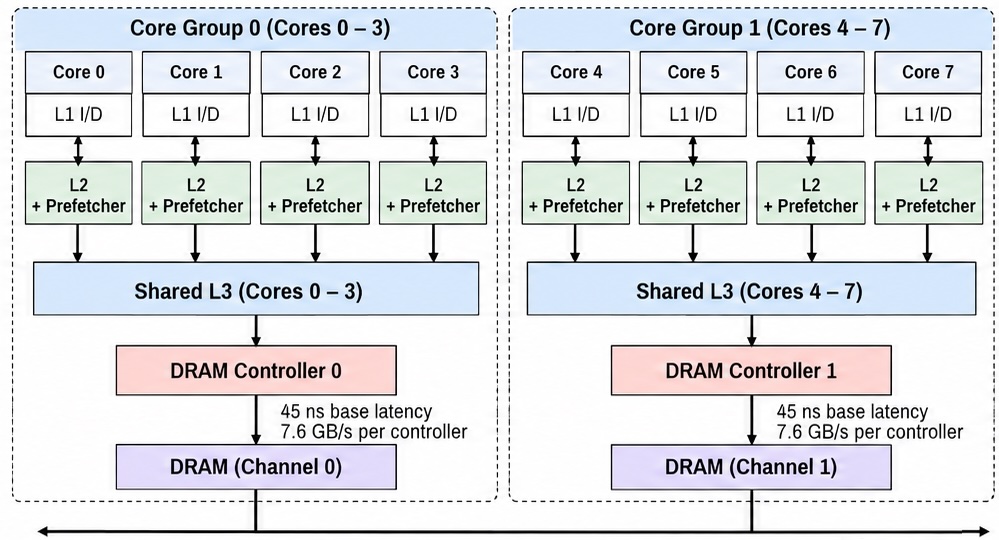}
    \caption{Linear Scaling System Architecture: Cache hierarchy and  sharing}
    \label{fig:LinScaleArch}
\end{figure}

\begin{table}[tb!]
\centering
\caption{Comparison of PF8 and D12/K1 Prefetcher, Runtime Improvement $\times$/\% less time}
\label{tab:pf8_d12k1_performance}
\renewcommand{\arraystretch}{1.25}
\setlength{\tabcolsep}{1pt}
\footnotesize
\begin{tabular}{|c|c|c|c|c|c|}
\hline
\textbf{Cores} & \textbf{PF8 } & \textbf{PF8} & 
\textbf{D12/K1} & \textbf{D12/K1} & \textbf{Runtime} \\
\textbf{} & \textbf{time} & \textbf{L2 miss} & 
\textbf{time} & \textbf{L2 miss} & \textbf{improvement} \\ 
\hline
1 &
41.206 ms &
60.09\% &
9.052 ms &
3.20\% & \cellcolor{yellow}
4.55$\times$ / 78.0\% \\
\hline
8 &
6.512 ms &
$\sim$60.14\% &
 4.462 ms &
 $\sim$3.30\% & \cellcolor{yellow}
1.46$\times$ / 31.5\%  \\
\hline
\end{tabular}
\end{table}

\subsection{Memory-Hierarchy Organization for SIMD Multicore}

\begin{figure*}[tb!]
    \centering
    \includegraphics[width=0.8\linewidth]{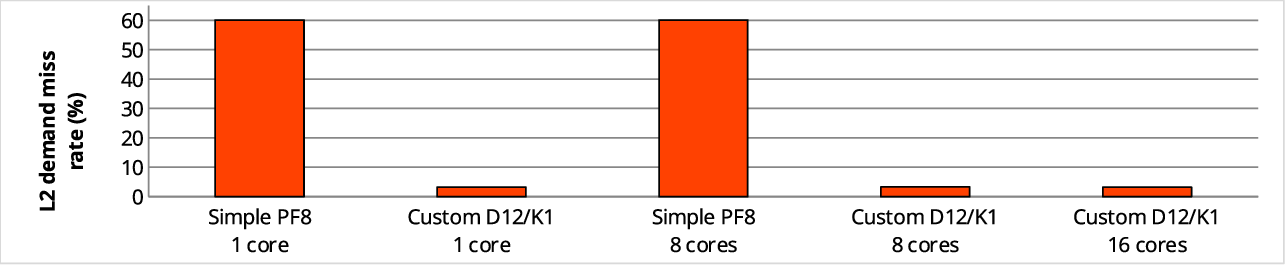}
    \caption{Conventional PF8 remains near 60\% L2 demand misses after AVX2, whereas D12/K1 stays near 3.2-3.4\% through 16 cores. 
    }
    \label{fig:L2_miss_rate}
\end{figure*}

As reported in Table \ref{tab:baseline_progression} and Table \ref{tab:lookahead_sensitivity}, a core configured with D12/K1 prefetch and AVX2 performs well in private L1 and private L2 configurations where L1 and L2 cache misses are around 3.2\%. We extended the same to multicore, considering the L2 private cache for each core and each L2 cache configured with D12/K1 prefetcher. 

With D12/K1, the L2 demand-miss rate remains approximately 3.2–3.4\% through 16 cores, indicating that most streamed weight lines are prefetched before they are requested by the processor. This suggests that the private L1 and L2 caches effectively absorb most of the memory access latency, while a shared L3 cache and DRAM organized in a scalable cluster-based manner can provide an effective architecture for sustaining performance as the core count increases.

Hence, we propose a memory hierarchy organization for the SIMD multicore architecture (termed LinScaleLLM), as illustrated in Figure~\ref{fig:LinScaleArch}, where the L1 and L2 caches are private to each core, while the L3 cache and DRAM controller/channel are shared among clusters of four processors. We conduct our experiments using this organization to evaluate the performance and scalability of SIMD-based LLM workloads.

Figure \ref{fig:L2_miss_rate} shows the miss rate single core and octa core system with AVX2. For conventional PF8 remains near 60\% L2 demand misses after AVX2, whereas D12/K1 stays near 3.2-3.4\% through 16 cores. Table~\ref{tab:pf8_d12k1_performance} compares the performance of the conventional PF8 prefetcher with the proposed D12/K1 configuration across 1 and 8 core executions. PF8 exhibits a high L2 demand-miss rate of approximately 60\% in both configurations, indicating that its fixed prefetch degree is unable to effectively hide memory access latency. In contrast, D12/K1 reduces the L2 demand-miss rate dramatically to 3.20\% and approximately 3.30\% for 1 and 8 core executions, respectively. This substantial reduction in cache misses translates directly into lower execution time. For a single core, D12/K1 reduces runtime from 41.206 ms to 9.052 ms, achieving a 4.55$\times$ speedup and 78.0\% reduction in execution time. With 8 cores, the runtime decreases from 6.512~ms to 4.462 ms, corresponding to a 1.46$\times$ speedup and 31.5\% reduction in execution time. These results demonstrate that decoupling the prefetch lookahead distance from the prefetch degree enables substantially more timely memory requests and improves the effectiveness of prefetching, particularly in the single-core case.

\begin{figure}[tb!]
    \centering          \includegraphics[width=0.51\textwidth]{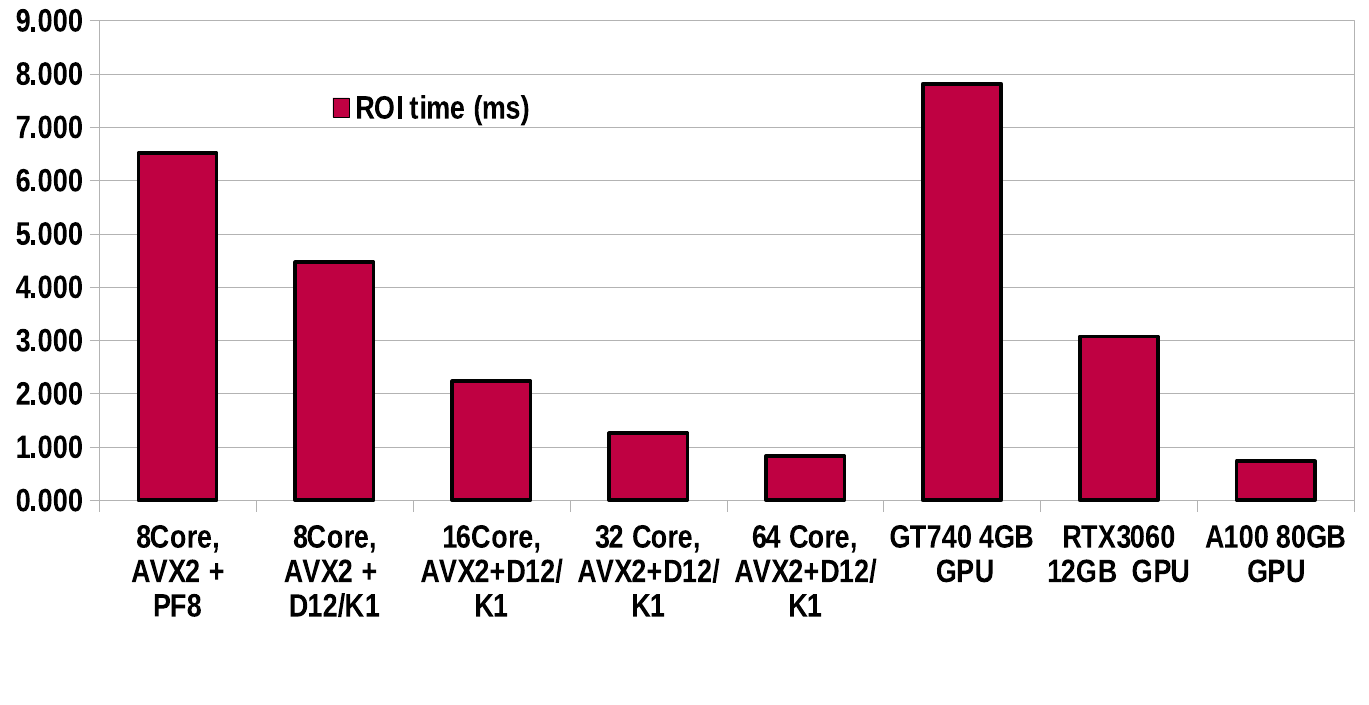}
  
    \caption{Performance evaluation of the proposed \ours~architecture and compared with low, medium and high end GPUs}
    
    \label{fig:performance_evaluation}
\end{figure}

\subsection{ Near-compulsory DRAM traffic and  Multicore scaling performance follows memory-controller count}

As shown in the Table \ref{tab:d12k1_scaling}, the 64 MiB matrix contains exactly 1,048,576 distinct 64-byte cache lines. Selected custom-prefetch runs generate total DRAM counts close to this floor: 1,049,782 at 1 core, about 1,050,022 at 8 cores, and 1,051,590 at 16 cores, and similar trend for 32-64 cores system. These are only about 0.12\%, 0.14\%, and 0.29\% above the matrix-line count, respectively. Thus, the large reduction in demand misses is not achieved by multiplying memory traffic; it primarily changes when the compulsory data is requested.

Traffic-efficiency can be interpreted in this nicely, as the custom prefetcher does not make the streaming matrix disappear from DRAM. It converts most first-touch matrix fetches from late demand misses into timely prefetches while keeping total traffic near the compulsory lower bound.

\begin{table}[tb!]
\centering
\caption{D12/K1 scaling using proposed hierarchy interm number of cores (NumDC: number of DRAM controller, Speedup is ratio of execution time on single core without prefetch and AVX to execution time  on  multicore with prefecth and AVX)}
\label{tab:d12k1_scaling}
\renewcommand{\arraystretch}{1.05}
\setlength{\tabcolsep}{2pt}

\begin{tabular}{|c|c|c|c|c|c|c|}
\hline
\textbf{Cores} & \textbf{Num} & \textbf{Time} &
\textbf{L2 } & \textbf{Total DRAM} & \textbf{DC} & \textbf{Speedup} \\
\textbf{} & \textbf{DC} & \textbf{(ms)} &
\textbf{miss} & \textbf{Access} & \textbf{BW util.} &  \\

\hline
1  & 1  & 9.052 & 3.20\% & 1,049,782 & 97.76\% & 11.71$\times$ \\
\hline
2  & 1  & 8.853 & 3.23\% & 1,049,826 & 99.95\% & 11.97$\times$ \\
\hline
4  & 1  & 8.868 & 3.25\% & 1,049,950 & 99.80\% & 11.96$\times$ \\
\hline
8  & 2  & 4.462 & 3.30\% & 1,050,022 & 99.19\% & 23.75$\times$ \\
\hline
16 & 4  & 2.233 & 3.22\% & 1,051,590 & 99.25\% & 47.47$\times$ \\
\hline
\rowcolor{yellow}
32 & 8  & \textbf{1.255} & 4.20\% & 1,051,590 & 98.72\% & \textbf{84.46$\times$} \\ 
\hline
\rowcolor{yellow}
64 & 16 & \textbf{0.831} & 5.28\% & 1,051,590 & 96.42\% & \textbf{127.56$\times$} \\ 
\hline
\end{tabular}
\end{table}

\subsection{Performance Comparison with GPU-based System}

We evaluate the MVM kernel on three NVIDIA GPUs spanning low range, mid range, and high-end configurations: (a) GT 740 with 4~GB VRAM, (b) RTX 3060 with 12~GB VRAM, and (c) A100 with 80~GB VRAM. Performance of the MVM kernel is reported along with the proposed system. Figure \ref{fig:performance_evaluation} shows execution time in milliseconds for the MVM kernel on different architecture configurations and GPU settings. At 64 cores, the CPU achieves MVM performance comparable to a high-end NVIDIA A100 GPU system. These numbers are 2.233, 1.255,  0.831 ms for 16-core, 32-core and 64-core CPU systems respectively (as reported in Table \ref{tab:d12k1_scaling}), and 7.81, 3.07 and 0.73 ms for GT740 4~GB, RTX3060 12~GB, A100 80~GB GPU system respectively. 

The reported CPU execution times are derived from the simulated cycle count within the Region of Interest (ROI). Since the experiments are conducted through cycle-level simulation, the simulated cycle count is converted into execution time using the processor's operating frequency, following the same methodology used by Sniper, which reports both execution time and the corresponding number of executed cycles. In contrast, the GPU execution time is obtained from an actual NVIDIA GPU system using NVIDIA's execution-time profiling tools. Therefore, CPU execution times represent \textbf{simulation-based estimates}, whereas GPU execution times are \textbf{measured directly from real hardware}.

\subsection{Complete LLM Decode Steps on \ours}
\label{sec:llm_decode}

The MVM kernel dominates decoding, but a complete decode step also runs attention over the cached keys and values,
normalization, the output layer and an OpenMP barrier after every operation. To check that the scaling of
Table~\ref{tab:d12k1_scaling} carries over, we simulate complete decode steps of llama.cpp~\cite{llamacpp} in the Sniper set-up of Table~\ref{tab:system_configuration} with the D12/K1 prefetcher. The model is
Llama-3.2-1B-Instruct~\cite{grattafiori2024llama3} in the Q8\_0 format (INT8 weights with one scale per 32 weights),
so each step streams 1.32~GB of weights. After a 7-token prompt and one warm-up step, the region of interest is
exactly one decode step, all 16 layers and the output layer, with $P$ threads on $P$ cores and cold caches. We build
llama.cpp with AVX2 and with AVX-512; as in Table~\ref{tab:d12k1_scaling}, the baseline is one core running llama.cpp
built without SIMD and without prefetching. Every simulated step generates the same token as the same build run
natively. With $N$ DRAM controllers, Sniper's set index leaves only every $N$-th set of a controller's directory
usable, so we give each directory $N\times$1,048,576 entries; each controller then tracks the 1,048,576 lines of the gainestown configuration.

\begin{table}[tb!]
\centering
\caption{Complete llama.cpp decode steps of Llama-3.2-1B (Q8\_0) on \ours~with D12/K1: time per generated token and
speedup  over one core without SIMD and without prefetching (1515.4~ms per token). DC BW util.: DRAM bandwidth used by the
AVX2 run, of 7.6~GB/s per controller.}
\label{tab:llm_decode}
\renewcommand{\arraystretch}{1.05}
\setlength{\tabcolsep}{1pt}
\begin{tabular}{|c|c|c|c|c|c|c|}
\hline
\textbf{Cores} & \textbf{Num} & \multicolumn{2}{c|}{\textbf{AVX2}} & \multicolumn{2}{c|}{\textbf{AVX-512}} & \textbf{DC BW} \\
\cline{3-6}
 & \textbf{DC} & \textbf{ms/token} & \textbf{Speedup} & \textbf{ms/token} & \textbf{Speedup} & \textbf{util.} \\
\hline
1 & 1 & 180.87 & 8.4$\times$ & 180.64 & 8.4$\times$ & 96.5\% \\ \hline
2 & 1 & 178.65 & 8.5$\times$ & 178.57 & 8.5$\times$ & 97.7\% \\ \hline
4 & 1 & 177.39 & 8.5$\times$ & 177.27 & 8.5$\times$ & 98.5\% \\ \hline
8 & 2 & 89.80 & 16.9$\times$ & 89.67 & 16.9$\times$ & 97.2\% \\ \hline
16 & 4 & 46.27 & 32.8$\times$ & 46.28 & 32.7$\times$ & 94.4\% \\ \hline
32 & 8 & 24.50 & 61.9$\times$ & 24.30 & 62.4$\times$ & 89.3\% \\ \hline
64 & 16 & 13.29 & 114.0$\times$ & 13.14 & 115.4$\times$ & 82.4\% \\ \hline
\end{tabular}
\end{table}

Table~\ref{tab:llm_decode} shows the result. On one core, AVX2 with D12/K1 generates a token 8.4$\times$ faster
than the baseline and already uses 96.5\% of the controller's bandwidth, so two and four cores sharing that
controller gain only 1.01$\times$ and 1.02$\times$. Each further cluster adds a controller, and every doubling of
the cores from 4 to 64 speeds up the step by 1.98$\times$, 1.94$\times$, 1.89$\times$ and 1.84$\times$: 64 cores generate a token in 13.29~ms (75~tokens/s),
114.0$\times$ faster than the baseline and 13.6$\times$ faster than one AVX2 core. The gain per doubling falls
slightly at 32 and 64 cores, where the controllers are less busy (89.3\% and 82.4\% of their bandwidth,
against 94.4\% at 16 cores) and the L2 demand-miss rate rises from 7.0\% at 16 cores to 10.4\% at 64.
Each step reads 1.002--1.008$\times$ the weight bytes from DRAM, i.e., near-compulsory traffic, as for the
kernel. AVX-512 changes the time per token by at most 1.2\%: once D12/K1 makes the step bandwidth-bound, wider
SIMD does not help, as Table~\ref{tab:avx2_avx512_mvm} shows for the kernel.

\section {MMM Performance on  LinScaleLLM Architecture}

In LLM  inference, matrix-matrix multiplication (MMM) arises naturally during the \emph{prefill} phase, where multiple input tokens
are processed simultaneously. The MMM operation is formulated as $X_{T \times H} W_{H \times O} = Y_{T \times O}$, where $X$ represents the input token activations, $W$ denotes the model weight
matrix, and $Y$ represents the resulting output activations. In our workload,
the hidden dimension is fixed at $H=4096$ and the output dimension is
$O=14336$. Therefore, the computation can be expressed as $ X_{T \times 4096} W_{4096 \times 14336}=Y_{T \times 14336}$. Here, $T$ denotes the number of LLM tokens processed together. We consider  four representative token dimensions: $T \in \{1,128,512,2048\}$. These configurations represent different LLM execution regimes, ranging from
single-token processing ($T$=1) to large token batches ($T$=2048) as given in Table \ref{tab:mmm_workload}.

To improve cache locality and exploit data reuse, the MMM computation is tiled
along all three dimensions. We use the following blocking factors:$B_T=32$, $B_H=512$, and $B_O=16$. Thus, the computation is decomposed into tiles of size $32\times512$ for the input activation matrix and $512\times16$ for the
weight matrix, producing a $32\times16$ output tile. For each output tile,
32 tokens are processed simultaneously, 512 elements of the hidden dimension
are consumed in each inner computation, and 16 output features are generated
together.

This blocking organization increases the temporal reuse of input activations
and weights while reducing data movement across the cache hierarchy and main
memory. It also provides appropriately sized computation units for exploiting
SIMD parallelism and multicore execution.

\begin{table}[tb!]
\centering
\footnotesize
\caption{MMM workload configurations}
\label{tab:mmm_workload}
\renewcommand{\arraystretch}{1.05}
\setlength{\tabcolsep}{3pt}
\begin{tabular}{|c|c|c|}
\hline
\textbf{$T$} & \textbf{LLM Execution Regime} & \textbf{MMM Shape} \\
\hline
1    & Single-token decoding / MVM-like & 
$1\times4096 \times 4096\times14336$ \\ \hline
128  & Small token batch / short prefill &
$128\times4096 \times 4096\times14336$ \\ \hline
512  & Medium token batch &
$512\times4096 \times 4096\times14336$ \\ \hline
2048 & Large token batch / long prefill &
$2048\times4096 \times 4096\times14336$ \\ 
\hline
\end{tabular}
\end{table}

As shown in Table \ref{tab:mmm_token_reuse}, for $T=1$, the weight matrix is accessed for a single token, making the
execution strongly influenced by memory bandwidth and weight movement, and
therefore similar to an MVM workload. As $T$ increases, the same weight
matrix is reused across multiple tokens, increasing arithmetic intensity and
providing greater opportunities for cache reuse, SIMD utilization, and
multicore parallelism. Consequently, varying $T$ from 1 to 2048 enables us to
study the transition from memory-intensive, MVM-like execution to
compute-efficient MMM execution in LLM workloads. The important result is not the absolute runtime, because these are sampled MMM experiments. When T=1, there is almost no cross-token weight reuse: IPC = 0.27 and L3 miss = 100\%. When T=128 and T=512, the same weight tile is reused across token rows: IPC rises to about 2.8 and L3 misses fall
dramatically.  When T=2048, the activation footprint reaches nominal LLC capacity and competing state begins producing capacity
pressure.

\begin{table}[tb!]
\centering
\renewcommand{\arraystretch}{1.05}
\caption{MMM token-reuse characterization, 1 core, no prefetch.}
\label{tab:mmm_token_reuse}
\setlength{\tabcolsep}{3pt}
\begin{tabular}{|c|c|c|c|c|c|}
\hline
\textbf{$T$} & \textbf{$X$ footprint} & \textbf{IPC} & \textbf{L2 miss} &
\textbf{L3 miss} & \textbf{Demand} \\
 &  &  &  &
& \textbf{DRAM} \\
\hline
1    & 4 KiB   & 0.27 & 84.86\% & 100.00\% & 1,025 \\ \hline
128  & 0.5 MiB & 2.77 & \cellcolor{yellow}7.82\%  & \cellcolor{yellow}8.45\%   & 1,026 \\ \hline
512  & 2 MiB   & 2.79 & \cellcolor{yellow} 7.27\%  & \cellcolor{yellow}\cellcolor{yellow} 3.25\%   & 1,496 \\ \hline
2048 & 8 MiB   & 2.28 & 7.15\%  & 33.07\%  & 71,649 \\
\hline
\end{tabular}
\end{table}

\begin{table}[tb!]
\centering
\caption{SIMD benefit and total speedup for reuse-rich $T=128$ MMM on multicore (D12/K1 prefetch and Memory Organization as shown in Figure \ref{fig:LinScaleArch}), SIMD speedup is ratio of scalar time to SIMD time, total speedup is ratio of single core scalar time to multicore SIMD time}
\label{tab:mmm_simd_benefit}
\setlength{\tabcolsep}{2pt}
\renewcommand{\arraystretch}{1.05}
\begin{tabular}{|c|c|c|c|c|}
\hline
\textbf{Cores} & \textbf{Scalar} &
\textbf{AVX2} & \textbf{SIMD} & \textbf{Total} \\ 
\textbf{} & \textbf{(ms)} &
\textbf{(ms)} & \textbf{Speedup} & \textbf{Speedup} \\ \hline
1 & 76.216 & 6.249 & \cellcolor{yellow} \textbf{12.20$\times$} & \cellcolor{yellow} \textbf{12.20$\times$} \\  \hline
2 & 38.109 & 3.301 & \cellcolor{yellow} \textbf{11.54$\times$} & \cellcolor{yellow} \textbf{23.09$\times$} \\ \hline
4 & 19.099 & 1.832 & \cellcolor{yellow} \textbf{10.42$\times$} & \cellcolor{yellow} \textbf{41.60$\times$} \\ \hline
8 & 9.593 & 1.070 & \cellcolor{yellow} \textbf{8.97$\times$} & \cellcolor{yellow} \textbf{71.23$\times$} \\
\hline
\end{tabular}
\end{table}

The results in Table \ref{tab:mmm_simd_benefit} demonstrate the substantial SIMD benefit of the reuse-rich $T=128$ MMM kernel across different core counts. With a single core, AVX2 reduces execution time from 76.216 ms to 6.249 ms, achieving a 12.20$\times$ speedup. As the number of cores increases, the SIMD speedup gradually decreases, from 11.54$\times$ at 2 cores to 10.42$\times$ at 4 cores and 8.97$\times$ at 8 cores. Overall speedup on 8 cores with SIMD is \textbf{71.23x} (which is the ratio of single-core scalar time to multicore SIMD time). Nevertheless, the AVX2 implementation continues to provide a large performance advantage over the scalar implementation. 
This behavior contrasts with low-reuse MVM, where SIMD acceleration is more strongly limited by streamed weight accesses and memory-delivery latency. In the $T=128$ MMM kernel, the reuse of matrix elements across multiple tokens allows the computational throughput provided by SIMD to translate much more effectively into execution-time reduction. The gradual reduction in SIMD speedup with increasing core count suggests increasing parallel resource contention, while the consistently large speedups indicate that the MMM kernel is able to exploit SIMD parallelism effectively due to its higher data reuse

\section{Conclusion and Future work}

In this work, we investigated the memory-hierarchy bottlenecks and multicore scalability of SIMDized LLM inference across both decode and prefill phases. We proposed \textit{LinScaleLLM}, a SIMD multicore architecture with private L1/L2 caches and clustered shared L3 cache and DRAM resources. Our characterization shows that SIMD acceleration significantly increases the weight-consumption rate of MVM, making private-cache access latency a primary bottleneck to multicore scalability. A lightweight D12/K1 L2 prefetching configuration improves prefetch timeliness while limiting additional DRAM traffic, effectively mitigating this bottleneck. With improved prefetch timeliness, memory bandwidth becomes the dominant constraint, highlighting the need to provision memory-controller resources alongside increasing core counts. For prefill, MMM benefits from substantial data reuse and compute intensity, enabling more effective utilization of SIMD resources compared with memory-bound decoding. Complete llama.cpp decode steps behave the same way: their time per token falls with the number of memory controllers, reaching 114.0$\times$ on 64 cores over one scalar core, and AVX-512 changes it by at most 1.2\% once the controllers are saturated. We demonstrate that scalable LLM inference on SIMD multicore processors requires coordinated design of the cache hierarchy, prefetching mechanisms, and memory-bandwidth resources.

We plan to extend \textit{LinScaleLLM} to larger and more diverse LLMs and investigate adaptive prefetching mechanisms that dynamically select the prefetch configuration at runtime based on the region of interest (ROI). We also plan to evaluate the proposed approach using low-level, industry-grade architectural simulators representative of modern processor designs to further validate its performance and scalability under realistic system configurations.

\bibliographystyle{alpha}

\end{document}